\documentclass[twocolumn,prd,superscriptaddress,nofootinbib]{revtex4-1}
\usepackage{natbib}
\usepackage[section]{placeins}

\usepackage{graphicx}
\usepackage{natbib}
\usepackage{multirow}
\usepackage[dvipsnames]{xcolor} 
\pdfoutput=1  

\usepackage{graphics,psfrag}
\usepackage{mathrsfs}

\newcommand{\nn}{\nonumber}

\def \be {\begin{equation}}
\def \ee {\end{equation}}
\def \bea {\begin{eqnarray}}
\def \eea {\end{eqnarray}}

\def \ble {\begin{widetext}\begin{equation}}
\def \ele {\end{equation}\end{widetext}}
\def \blea {\begin{widetext}\begin{eqnarray}}
\def \elea {\end{eqnarray}\end{widetext}}

\def \nn {\nonumber}

\newcommand{\eq}[1]{(\ref{#1})}

\def \Re {{\textrm{Re}}}

\def \and {{\textrm{and}}}

\begin{document}
\title{Deep Generative Models of Gravitational Waveforms via Conditional Autoencoder}

\author{Chung-Hao Liao}
\email{rossliao125@gmail.com}
\affiliation{Department of Physics, National Taiwan Normal University, Taipei 11677, Taiwan}

\author{Feng-Li Lin}
\email{fengli.lin@gmail.com, corresponding author.}
\affiliation{Department of Physics, National Taiwan Normal University, Taipei 11677, Taiwan}
\affiliation{Center of Astronomy and Gravitation, National Taiwan Normal University, Taipei 11677, Taiwan}

\date{\today}

\begin{abstract}
We construct few deep generative models of gravitational waveforms based on the semi-supervising scheme of conditional autoencoders and their variational extensions. Once the training is done, we find that our best waveform model can generate the inspiral-merger waveforms of binary black hole coalescence with more than $97\%$ average overlap matched filtering accuracy for the mass ratio between $1$ and $10$. Besides, the generation time of a single waveform takes about one millisecond, which is about $10$ to $100$ times faster than the EOBNR algorithm running on the same computing facility. Moreover, these models can also help to explore the space of waveforms. That is, with mainly the low-mass-ratio training set, the resultant trained model is capable of generating large amount of accurate high-mass-ratio waveforms. This result implies that our generative model can speed up the waveform generation for the low latency search of gravitational wave events. With the improvement of the accuracy in future work, the generative waveform model may also help to speed up the parameter estimation and can assist the numerical relativity in generating the waveforms of higher mass ratio by progressively self-training.
\end{abstract}
 
\maketitle

\section{Introduction}\label{Introduction}

LIGO/Virgo has detected about a hundred of compact binary coalescence (CBC) up to its O3 observations \cite{Abbott:2016blz,LIGOScientific:2018mvr,Abbott:2020niy}. This is remarkable achievement of modern science. Due to the limitation of LIGO/Virgo's sensitivity, these events are detected by the method of matched filtering \cite{schutz_1991,Owen_1996,Owen_1999}, which calculates the overlap between the whitening data and the theoretical gravitational waveform templates.  Similarly, the source properties of these events are also extracted based on matched filtering to perform the  Markov-Chain-Monte-Carlo (MCMC) Bayesian parameter estimation (PE) \cite{veitch2015parameter,biwer2019pycbc}. In both processes of detection and PE of gravitational wave events, a huge number of theoretical waveform templates are required for matched filtering, therefore the efficiency of evaluating waveform templates is crucial for detection and to accelerate the PE procedures. However, due to the nonlinear feature of Einstein gravity and the unavoidable strong gravity regime for the mergers of two compact objects, it is notoriously difficult to calculate the CBC dynamics and the associated gravitational waveforms. For example, it is known \cite{Hinder:2010vn,Pfeiffer_2012} to require about hundred thousand CPU hours to obtain a state-of-art CBC waveform by solving numerical relativity. The required computing time will be increased by one or two order for the higher mass-ratio CBC events, and is beyond what the current computing facility can afford. Thus, it is impractical to adopt such ab initio waveforms directly for performing either detection or PE.

To accelerate the generation of theoretical waveforms for practical applications, some analytical waveform models are introduced with a few parameters to be fitted by the results of numerical relativity. The well-known examples are IMRPhenomP models \cite{Hannam:2013oca,Khan:2018fmp}, the synergy models \cite{Buonanno_2007,Pan_2011,Ajith_2011} by combining the post-Newtonian \cite{Einstein:1938yz,Blanchet:2013haa},  effective-one-body (EOB) formalism \cite{PhysRevD.59.084006,Buonanno_2000}, black hole perturbation \cite{Kokkotas_1999,Nollert:1999ji} and numerical relativity \cite{Centrella:2010mx,Loffler:2011ay}, and the reduced order models or surrogate models \cite{Blackman_2015,P_rrer_2016,Williams_2020} which span the generic waveforms with some orthonormal basis. 
However, it still takes few hundredths to few tenths of a second to evaluate a single waveform based on  the aforementioned analytical waveform models \footnote{This can be estimated by generating the waveforms from template library in either PyCBC or GstLAL.}.  By this speed of waveform generation, it will usually take weeks or even months to obtain the state-of-art PE results for a single event based on the MCMC algorithm. One can then expect the overall computing power cost or time-span for PE will increase rapidly for the latter operations of LIGO/Virgo/KAGRA such as O4 or O5, for which the number of detection CBC events will be increased by an order or more. Therefore, the speed-up of waveform generation becomes a pressing issue even in the near future.

Besides, those aforementioned analytical waveform models are in nature interpolating models by fitting the parameters with a known set of waveforms. This implies that the model could become more complicated and cumbersome when the range of the waveforms are extended, such as going to higher mass ratio. The increase of the complexity will reduce the models' efficiency of generating the real-time waveforms for the detection or PE. Thus, it is crucial to have some extrapolating models of waveform generation to resolve the conflict between complexity and efficiency of the traditional analytical waveform models. 
  
\begin{figure}[htpb]
\resizebox{6.5cm}{!}{\includegraphics{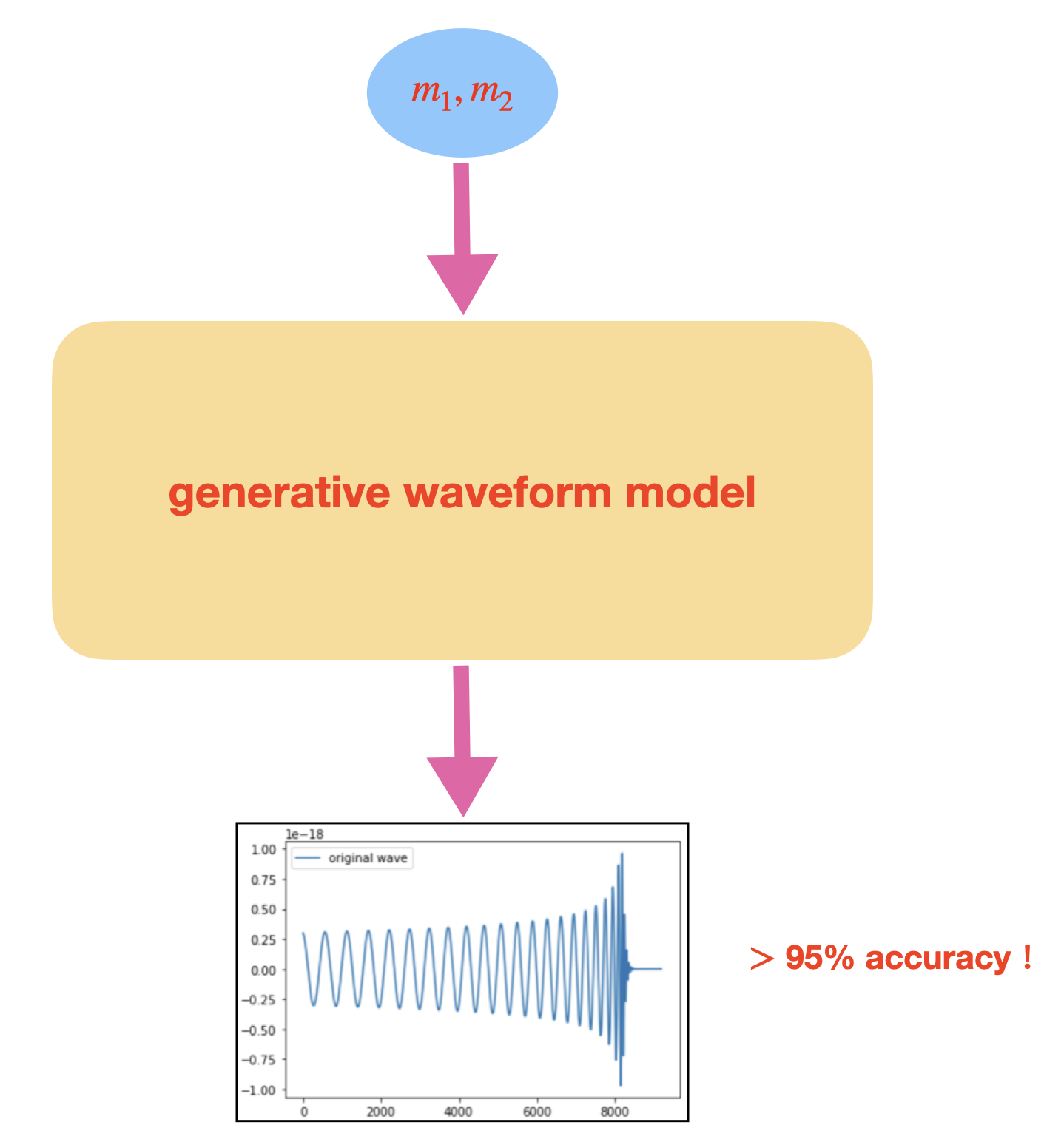}}
\caption{A generative model of gravitational waveforms: Once the neural network model is well-trained, it can generate the gravitational waveforms with more than $95\%$ accuracy when providing just the source labels such masses $(m_1,m_2)$ of the binary black holes. The accuracy rate of the waveforms is defined in \eq{accuracy-rate} below. As a preliminary study for the proof-of-concept, in this work we mainly consider the inspiral-merger parts of the full waveforms.}
\label{generative model}
\end{figure}

Motivated by the above discussions on the limitations of the known models of waveform generation, we turns to the deep learning for the resolution. We aim to construct some deep learning neural network to generate the CBC gravitational waveforms of high accuracy by giving the source parameters such as the masses, spins of the binary compact objects, as schematically depicted in Fig. \ref{generative model}. Even the training time will be increased as the training set is enlarged, the time of evaluating a new waveform with the trained machine will not be increased much. This then resolves the conflict between the complexity and efficiency for the real-time applications.

Moreover, we also hope this deep learning neural network to be generative so that it can generate the waveforms which do not belong to the source parameter ranges of the training set. For example, we can train the machine with waveforms of only low mass ratio (LMR), and then generate the accurate waveforms of higher mass ratio (HMR). This could help to efficiently obtain the HMR waveforms, which will be computationally costly by numerical relativity. 
However, in this work we will not explore this scenario but just a toy version, for which we employ the training set containing small fraction of HMR waveforms to demonstrate the possibility.

In view of the above target features, this deep learning machine should be the supervised one when training with the given source parameters and the associated waveforms. On the other hand, it is also better to be generative and the unsupervised one so that it has the potential to turn into an extrapolating model of generating the HMR waveforms.  For this purpose, in this paper we adopt the conditional (variational) autoencoder (cAE or cVAE) \cite{Sohn2015LearningSO,nguyen2017plug,tonolini2020variational} to construct various deep learning models to generate CBC gravitational waveforms \footnote{The cVAE framework is recently adopted as the generative model of posteriors for the PE of the CBC events, see \cite{gabbard2019bayesian,Green:2020hst,green2020complete}.}. This scheme belongs to the so-called semi-supervised learning by combining both features of supervised and unsupervised learning \footnote{For the basic discussion of supervised and unsupervised learning, please see \cite{GoodBengCour16,Carleo_2019}.}. It is built on a more basic scheme for the unsupervised learning, the autoencoder (AE) \cite{10.5555/2987189.2987190}, or its generative extension, the variational autoencoder (VAE) \cite{kingma2019introduction,yu2020tutorial}. We will introduce the basics of these neural networks in the next section.

\begin{figure}[htpb]
\resizebox{\hsize}{!}{\includegraphics{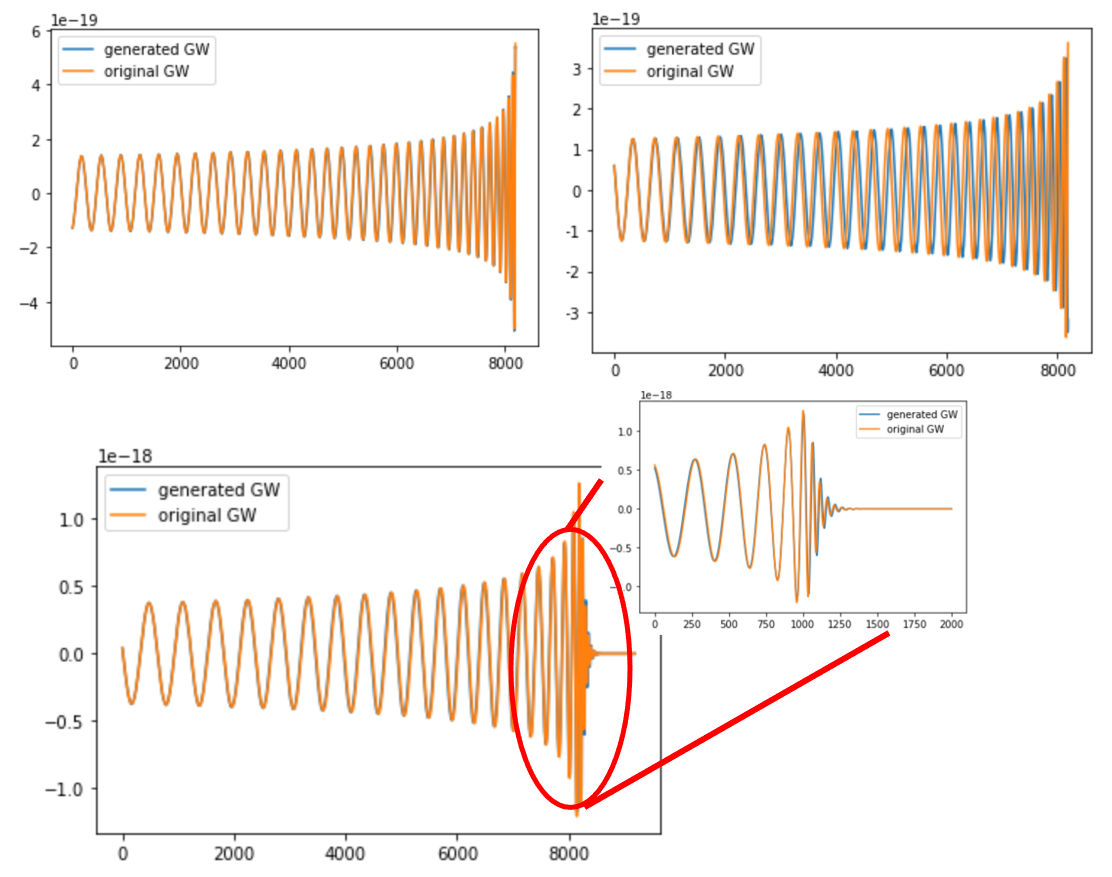}}
\caption{Some typical examples of the generated waveforms by well-trained cVAE models with different accuracy rates. {\bf Top-left:} a generated inspiral-merger waveform of $99.50\%$ accuracy when comparing with the corresponding EOB waveform by overlap match. {\bf Top-right:} a generated inspiral-merger waveform of $92.37\%$ accuracy. {\bf Bottom:} A generated full inspiral-merger-ringdown waveform of $99.74\%$ accuracy.}
\label{Typical example}
\end{figure}

As a preliminary study for the proof-of-concept, in this work we mainly consider the inspiral-merger parts of the full waveforms but truncating the ringdown part. We find that our best generative models can produce the waveforms with accuracy higher than $97\%$ even for the generation of HMR waveforms. Moreover, it can produce a single waveform within one millisecond, which is about 10 to 100 times faster than producing an EOB waveform on the same computing facility. To visualize the accuracy rate, in Fig. \ref{Typical example} we shows some typical examples of the waveforms with different accuracy rates. 

The rest of the paper is organized as follows.  In the next section, we will briefly sketch the basics of autoencoder and its extensions including VAE and the conditional versions. In section \ref{sec: waveform data} we describe the tomography of our training data set, and how we prepare our training waveforms. Besides, the fitting factor or faithfulness (FF) based on the overlap of matched filtering is introduced to characterize the accuracy of the generative waveform models. In section \ref{schemes} we consider four waveform models based on the cAE scheme, and then summarize their accuracy and run-time in Table \ref{summary acc 20} and \ref{summary speed 20}, respectively. By comparing the accuracy, we pick up the best cAE waveform model and present its detailed information.  Finally, we conclude this paper in section \ref{conclusion}. In the Appendix, we present the performance of the cVAE counterparts of the cAE waveform models considered in the main text.

\section{Autoencoder and its extensions}\label{section:cVAE}

\begin{figure}[htpb]
\resizebox{4.5cm}{!}{\includegraphics{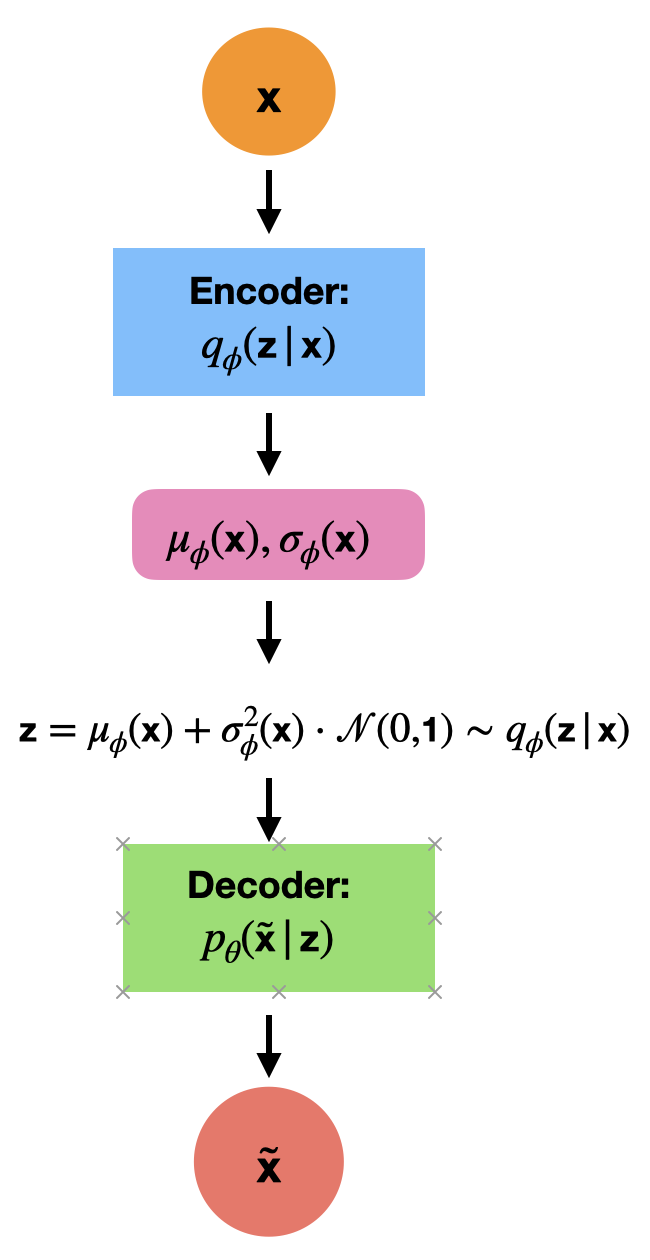}}
\caption{Schematic structure of a AE or VAE. It contains two components. (i) An encoder $q_{\phi}(z|x)$ which transforms an input vector $x$ to a latent vector $z$, which is deterministic for AE, but stochastic for VAE, i.e., $z=\mu_{\phi}(x)+\sigma^2_{\phi}(x) {\cal N}(0,1)$. Here ${\cal N}(0,1)$ is the unit normal distribution. (ii) An decoder $p_{\theta}(\tilde{x}|z)$ which transforms $x$ to an output $\tilde{x}$. The loss function of AE is just the reconstruction loss such as mean-squared error (MSE) between $\tilde{x}$ and $x$. On the other hand, the loss function of VAE contains two parts: the reconstruction loss and the Kullbac-Leibler (KL) loss as discussed in \eq{VAE loss}.}
\label{VAE}
\end{figure}

Our goal is to construct some deep learning models of gravitational waveforms as depicted in Fig. \ref{generative model}. The basic structure of this generative model is the so-called autoencoder (AE) \cite{10.5555/2987189.2987190} or it extension, the variational auto encoder (VAE) \cite{kingma2019introduction,yu2020tutorial}. The basic structure of AE and VAE is shown in Fig. \ref{VAE}, which contains two parts: the encoder and the decoder. The encoder (denoted by $q_{\phi}(z|x)$ with $\phi$ the abbreviation of biases and weights of the encoder's neural network) compress the input data $x$ into the latent layer $z$ of smaller dimensions than the ones of $x$, and then the decoder (denoted by $p_{\theta}(\tilde{x}|z)$ with $\theta$ the abbreviation of biases and weights of the decoder's neural network) un-compress the latent layer back to the final result $\tilde{x}$ of the same dimension as $x$. One then use some distance measure such as mean-squared error (MSE) between $x$ and $\tilde{x}$ as the reconstruction loss. The goal is to minimize the reconstruction loss to optimize the biases and weights of the whole AE's neural network. Since there is no label for the input data, this is the unsupervised learning.

Since the AE is a deterministic machine so that it may lack the power of extrapolations and could fail to be generative. To remedy this drawback, the VAE is introduced by making the latent layer a stochastic one. This is done by generating the means and variances of the Gaussian distributions as the output of the encoder, from which one can sample a latent layer as the input to the decoder, as shown in the middle of Fig. \ref{VAE}. The uncertainty of the layer make the VAE to be able to ``think out of the box", thus a generative machine. However, besides the reconstruction loss one should also consider the regularization loss which characterizes how much the stochastic latent layer deviates from ${\cal N}(0,1)$, i.e., the unit Gaussian with zero mean. This is measured by their Kullbac-Leibler (KL) divergence. It turns out that the combined loss is equal to upper bound of the negative of the log likelihood of the input data distribution $p_{\theta}(x)$, i.e., 
\bea\label{VAE loss}
 -\log p_{\theta}(x)&\le &{\bf E}_{z\sim q_{\phi}(z|x)} [-\log p_{\theta}(\tilde{x}|z)] \qquad
 \nn\\ \label{VAE-loss}
 &+& {\bf D}_{KL}[q_{\phi}(z|x)||{\cal N}(0,1)] \qquad 
\eea
where the first term on the right-hand-side is the reconstruction loss and the second term is the regularization loss.

\begin{figure}[htpb]
\resizebox{7cm}{!}{\includegraphics{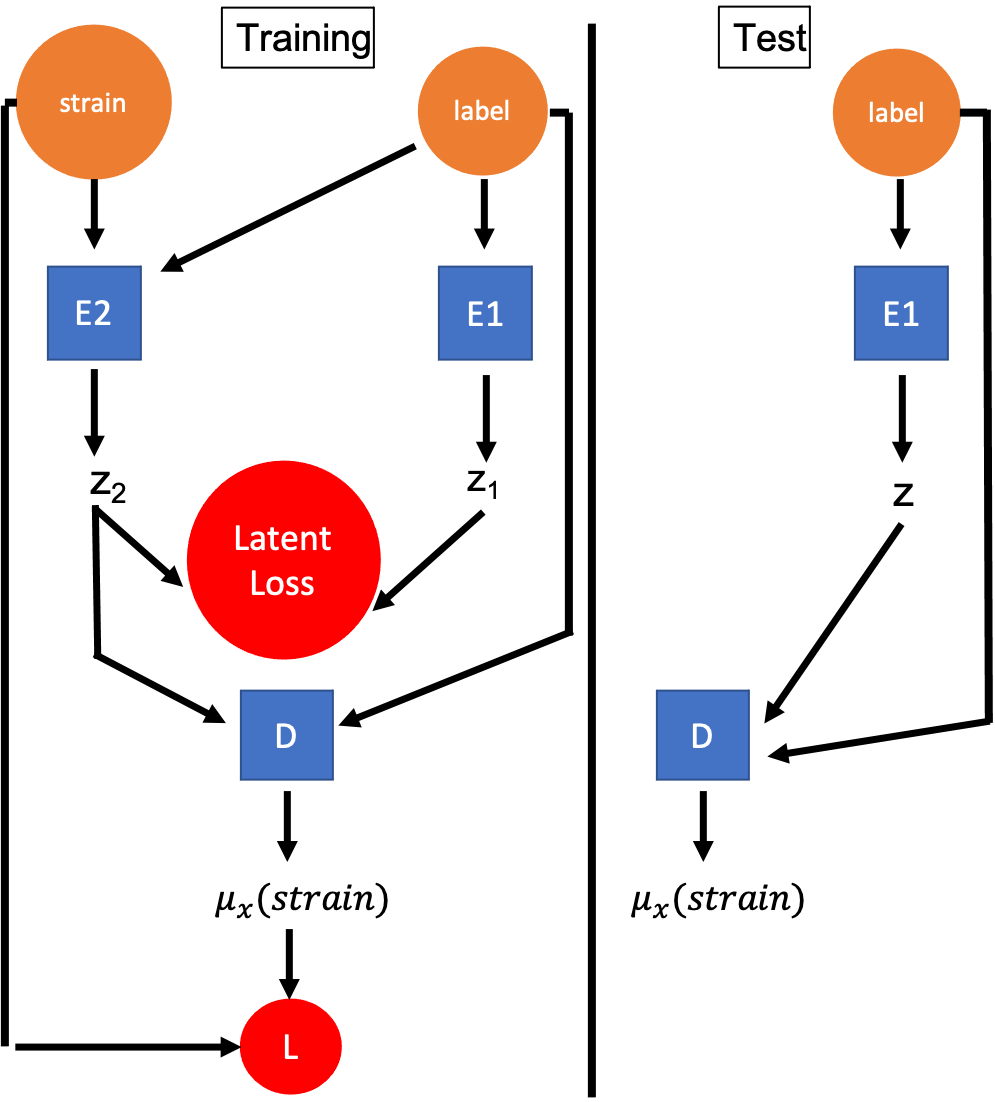}}
\caption{The schematic structure of a cAE or cVAE as a generative waveform model. {\bf Left panel:}
During the training period, it needs two encoders: (a) one for training the input data such as strains/waveforms, and (b) one for training the source labels associated with the input data such as $(m_1,m_2)$. {\bf Right panel:} After the training, the encoder (a) is removed so that it becomes a generative model, namely it generates waveforms by providing only the associated source labels.}
\label{cAE}
\end{figure}

When training the waveform models, the input $x$ from the training data set is the theoretical waveform such as EOB waveform, and we call it strain for short. We can choose the reconstruction loss to be the MSE between $x$ and $\tilde{x}$. The training process is to optimize the biases and weights of the neural network by minimizing the reconstruction loss such that the generated $\tilde{x}$ can be as close to $x$ as possible. After the training, the decoder can be turned into a generative model of strains, namely, given some input latent vector, the decoder will output some strain. However, this machine is not so useful in generating the strains with specific source properties because the latent space may not correspond to the required parameter space of the physical source properties, such as masses $(m_1,m_2)$ of the binary black holes. For convenience we call the source parameters the labels. To make the AE or VAE useful for our purpose, we adopt the way of semi-supervised training by also conditioning the labels when training the machine. After the training we will truncate the encoder part associated with the strain input, the remaining one with the label as input will then become the useful generative model of strains, namely, given the label such as $(m_1,m_2)$, the machine will generate the associated strain. The above scheme is called the conditional AE/VAE abbreviated as cAE/cVAE \cite{Sohn2015LearningSO,nguyen2017plug,tonolini2020variational}, and the basic structure is depicted in Fig. \ref{cAE} where an additional encoder for the labels of input data is introduced.

Due to the additional encoder, we now have two latent vectors $z_1$ and $z_2$ as shown in Fig. \ref{cAE}. We can then introduce the latent loss to measure their difference. For AE, the latent loss can be MSE between latent vectors, but for VAE it is the KL divergence between the Gaussian distributions generated by the two encoders. On the other hand, the reconstruction loss is the MSE for both AE or VAE.

\section{Waveform data preparation and overlap accuracy}\label{sec: waveform data}

\begin{figure}[htpb]
\resizebox{\hsize}{!}{\includegraphics{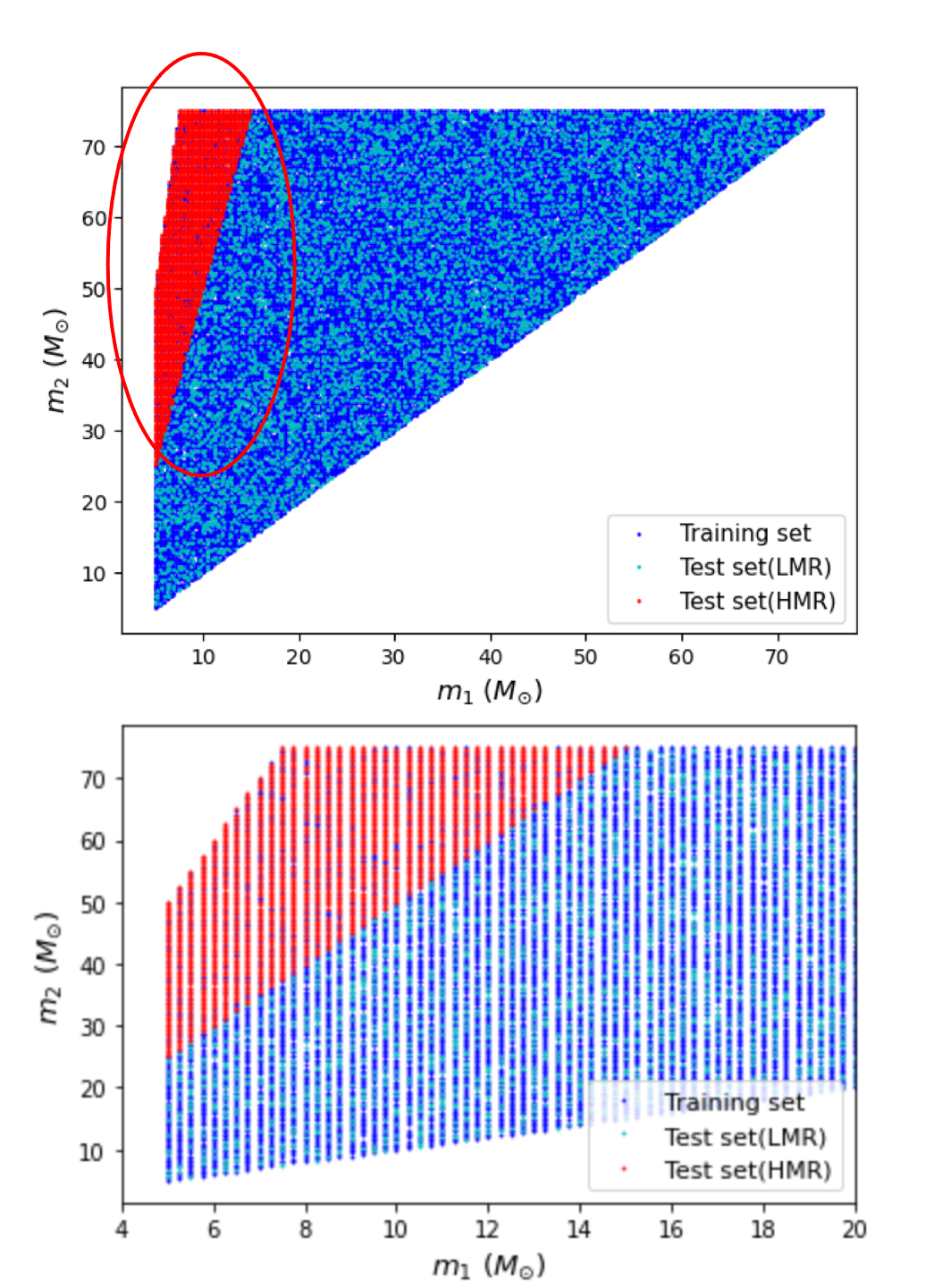}}
\caption{Tomography of data set for training, validation and test of a cAE model with the numeric as listed in Table \ref{Data Set}. {\bf Top:} Overview of the tomography. {\bf Bottom:} Enlarged view of some portion circled in the Top figure, with more clear visual estimate. }
\label{tomography}
\end{figure}

Once we construct the code for cAE or cVAE, we prepare a set of strains to train the machine. A strain is the linear combination of the two polarization modes $h_+(t)$ and $h_{\times}(t)$, i.e.,
\be\label{strain}
h(t)=h_{+}(t)+i h_{\times}(t)\;.
\ee
To be specific, we consider the inspiral-merger part of the CBC strains of binary black holes with their masses $(m_1,m_2)$ as the only source parameters, i.e., without spin and procession. We further divide the set into two subsets, one is called the low-mass-ratio (LMR) set for $q=m_2/m_1\le 5$ and the other one called the high-mass-ratio (HMR) set for $q>5$. These strains are obtained from ENOBNR \cite{Buonanno_2007} of the PyCBC library \cite{biwer2019pycbc}, in which each strain is divided into $8192$ time segments. We basically train the machine mainly with the LMR set but combining with about $20\%$ of HMR strains. The latter is used as the tutor seed to train the machine toward a generative model for the other $80\%$ of HMR strains. The tomography of our training and test sets is shown in Fig. \ref{tomography}, and in Table \ref{Data Set} we give the more details for the specs of this tomography.   Moreover, the fraction of the HMR templates in the total training set is only about $2.46\%$. This tiny fraction is chosen on purpose to mimic as closely as possible the real extrapolating model for which the training set contains no HMR waveform.

\begin{table}[htpb]
\centering
\begin{tabular}{|*{8}{c|}}
\hline
{} & $m_1$ & $m_2$ & $q$ & $\Delta m$ & train & valid & test \\
\hline
$q\le 5$ & [5.0,75.0] & [5.0,75.0] & [1,5] & 0.25 &  24865  & 3552 & 7104  \\
\hline
$q > 5$ & [5.0,75.0] & [5.0,75.0] & [5,10] & 0.25 & 682  & 36  & 2873  \\
\hline	 
\end{tabular}
\caption{The range of source parameters and the amounts of the data set. Here the mass ratio is denoted by $q\equiv m_2/ m_1$ with $1\le q\le 10$. The corresponding percentages of (training, validation, test) is (70\%, 10\%, 20\%) for LMR (low-mass-ratio) data ($q\le 5)$, and is (19\%, 1\%, 80\%) for HMR (high-mass-ratio) data ($q>5$). Note that the fraction of the HMR templates is only about  $2.46\%$ of the total training data set, including both training and validation data.}
\label{Data Set}
\end{table}

\begin{figure}[htpb]
\resizebox{7.5cm}{!}{\includegraphics{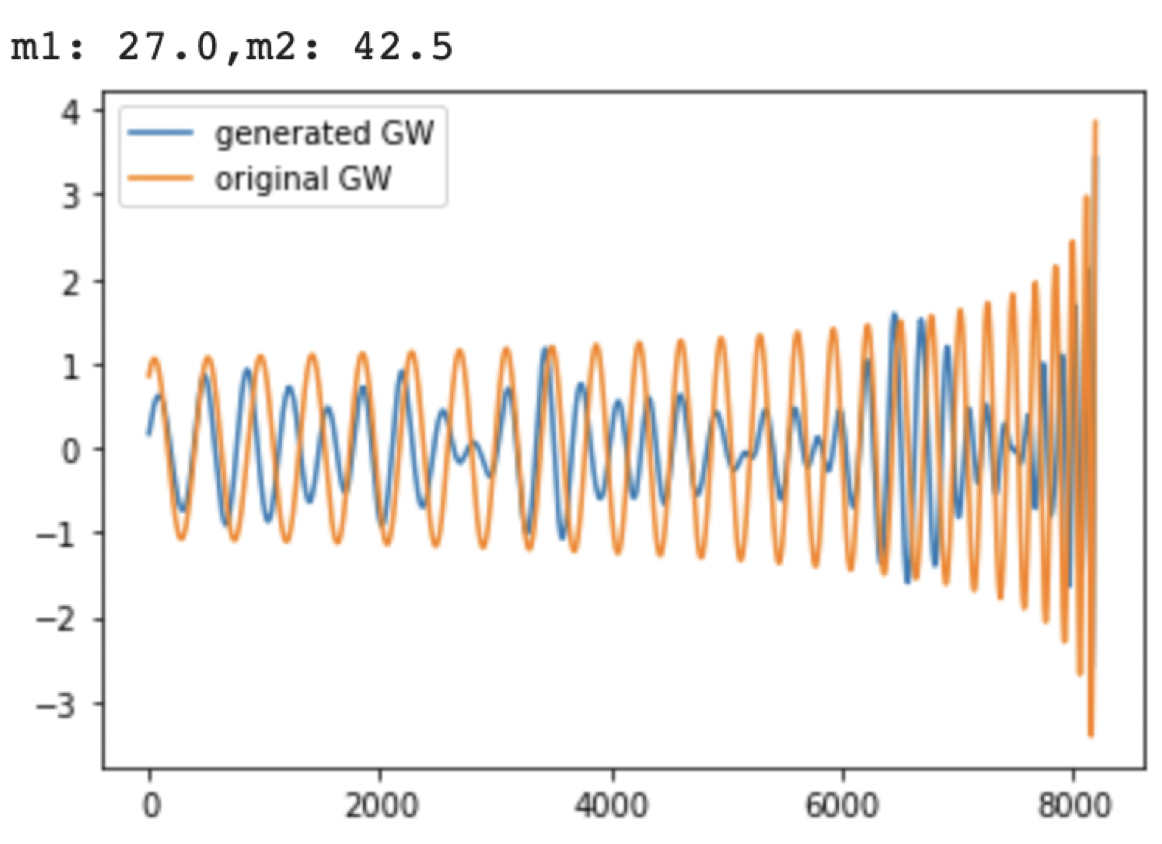}}
\caption{A typical generative waveform (blue color) from a trained cAE or cVAE by inputting a inspiral-merger parts of the CBC strains (orange color) in a time series format. It cannot catch both the phase and amplitude at the same time.}
\label{fail-1}
\end{figure}  

As a preliminary study to demonstrate that a generative model of gravitational waveform is in principle possible, we will not consider the full CBC strain but truncate the ringdown part which is far shorter than the other part of the strain. The truncated waveform is denoted as the inspiral-merger strain. The purpose of this truncation is to further reduce the complexity of the frequency/amplitude part of a strain caused by the sudden change at the merger, and will help to well train the machine with less efforts in tuning the hyper-parameters.

\begin{figure}[htpb]
\resizebox{\hsize}{!}{\includegraphics{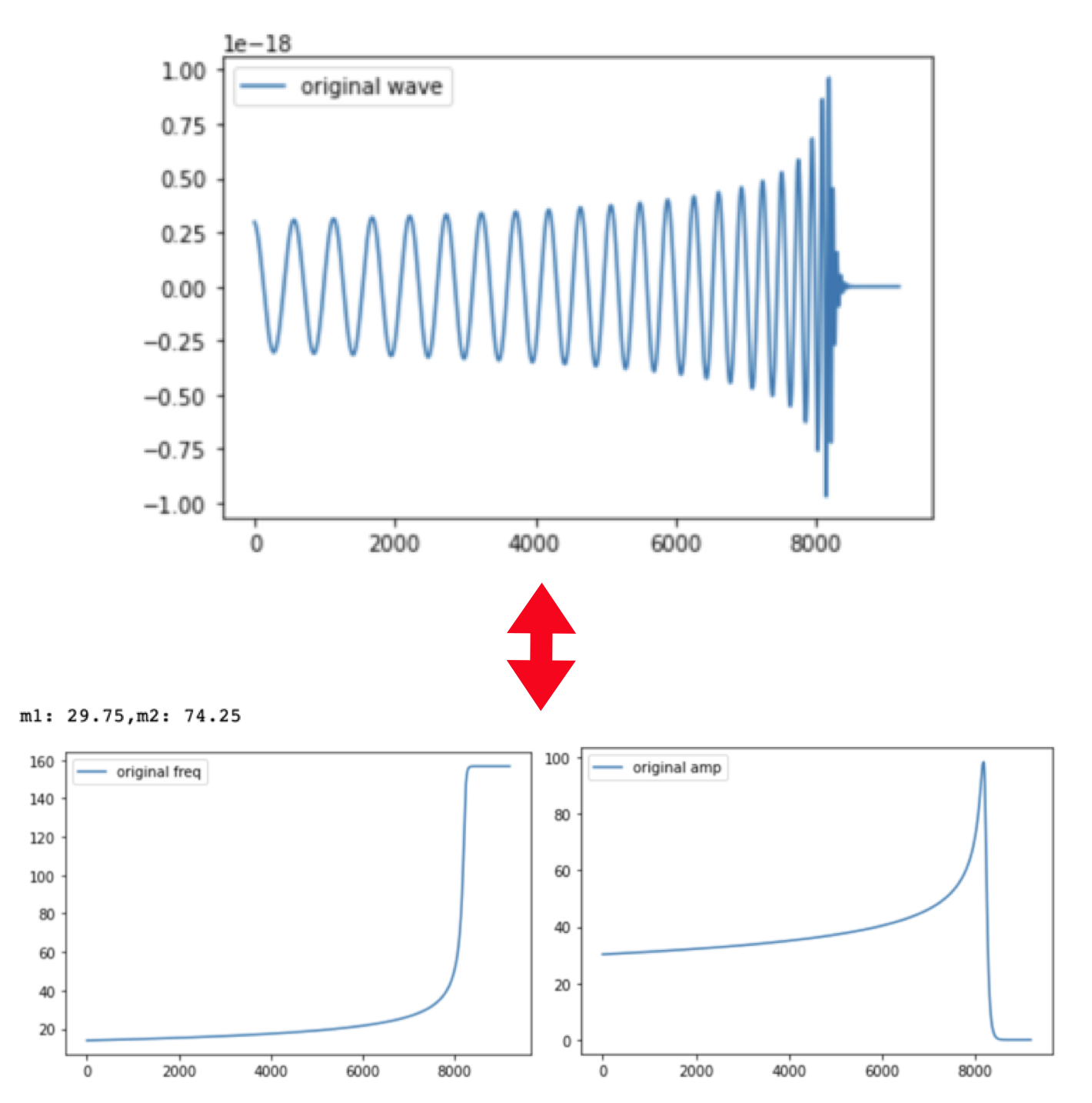}}
\caption{Decomposition of a time series strain $h(t)$ (top) into frequency (bottom-left) and amplitude (bottom-right) by using \eq{frequency vs t} and \eq{amplitude vs t}. The amplitude part has been multiplied by a factor of $10^{20}$ to have comparable scale with the frequency one.}
\label{decomposition}
\end{figure}

Using the time series form of the inspiral-merger strains to train a cAE or cVAE, the result turns out to be not good for reasonable machine size and training, see Fig. \ref{fail-1} for a typical result. It implies that the model cannot catch up the amplitude and phase correctly at the same time. This suggests that this form of strain is still too complicated for a cAE or cVAE of reasonable size to work properly. Motivated by this result, we then decide to separate the amplitude and frequency parts of a strain, and then juxtapose them as the input of the cAE or cVAE. To be specific, from the two polarization modes we first obtain the instantaneous phase
\be
\theta(t)=\tan^{-1}\Big({h_{\times}(t) \over h_+(t)}\Big)
\ee
and the instantaneous frequency and amplitude are given respectively by
\bea
&& \omega(t+{\delta t \over 2})={\theta(t+\delta t) -\theta(t) \over 2\pi \delta t} \label{frequency vs t}
\\
&& A(t)=\sqrt{h^2_{+}(t)+h^2_{\times}(t)}. \label{amplitude vs t}
\eea
A typical example showing the above decomposition is given in Fig. \ref{decomposition}.

Even using this frequency/amplitude separated form of the strains to train the cAE or cVAE model, the result is still not good because the magnitudes of the input data have not been rescaled to avoid too small or too large values. This however can be solved as in the usual deep learning process for neural network by just normalizing the input data \cite{franoischollet2017learning}. The way of normalization we adopt is as follows:
\be
\hat{\omega}(t)={\omega(t)-\mu_{\omega} \over \sigma_{\omega}}\;, \qquad \hat{A}(t)={A(t)-\mu_A \over \sigma_A} \label{normalization}
\ee
where the normalization parameters $(\mu_{\omega},\sigma_{\omega})$ are respectively mean and variance evaluated from the 8192 segments of $\omega(t)$ \footnote{Due to the nature of its definition from the difference between two neighbor segments, there are only $8191$ segments for $\omega(t)$.}, and similarly for $(\mu_A,\sigma_A)$. We call these four parameters the key (to reconstruct the associated un-normalized strain).

Naively, we can juxtapose these four normalization parameters with the normalized strain vector $(\hat{\omega}(t), \hat{A}(t))$ of $2\times 8192$ segments to form the input of cAE or cVAE. However, their dimensions are not in proportion, the juxtaposition could suppress the significance of the key during the training, which will induce the unbearable error in recovering the full strain via \eq{normalization}. Therefore, we need to find the appropriate cAE or cVAE schemes to well train both the normalizing strains and the associated keys to the desirable accuracy. 

Once the training of a waveform deep learning model is done, we need to evaluate their performance based on some criterion of accuracy by comparing a machine-generated waveforms $h_{\rm ML}(t)$ with the corresponding waveform $h_{\rm EOB}(t)$ obtained from EOBNR. To calculate the accuracy we adopt the conventional overlap method used in gravitational waveform community. The overlap method is motivated by the matched filtering \cite{schutz_1991,Owen_1996,Owen_1999} for the signal detection or parameter estimation, in which the overlap between two waveforms $h_1(t)$ and $h_2(t)$ is defined by
\be
\langle h_1|h_2 \rangle=4 \; \Re \int_0^{\infty} \frac{\tilde{h}_1(f) \tilde{h}_2(f)^*}{S_n(f)} df
\ee
where $\tilde{h}_i(f)$ is the Fourier transform of $h_i(t)$ and $S_n(f)$ is the power spectral density (PSD) of detector's noise. In practical, some appropriate low and high frequency cutoffs will be imposed when performing the integral. To evaluate the accuracy of a waveform model, the following fitting factor (FF) or faithfulness \cite{Babak:2006uv,Buonanno_2007,Williams_2020} is adopted to compare $h_{\rm ML}(t)$ generated by our waveform model and the standard EOB waveform $h_{\rm EOB}(t)$,
\be\label{accuracy-rate}
 \textrm{FF}= \max_{t_0,\phi_0}\left[ \frac{\langle h_{\rm EOB}|h_{\rm ML}\rangle}{\sqrt{\langle h_{\rm EOB}|h_{\rm EOB}\rangle \langle h_{\rm ML}|h_{\rm ML}\rangle}}\right]
\ee
where $t_0$ and $\phi_0$ are respectively the initial time and inital phase of $h_{\rm EOB}(t)$.  Without being biased by the detector noise, below we will choose flat PSD, i.e., $S_n(f)=1$ for evaluating the FF \cite{Williams_2020}. To characterize the performance, we need to evaluate the FF of each template in the test data set, i.e., $20\%$ of LMR and $80\%$ of HMR, and find out the distribution of FFs, which can also be represented by its maximum, median and minimum. However, for simplicity we can represent and denote the accuracy simply by the average of the FFs over the test data set. This may not be precise enough but is more convenient when comparing the performances of different generative waveform model. Later on, we will give the cumulative distribution function of FFs and the associated maximum, median and minimum for the best model selected by comparing the average of FFs over the test data set. Moreover, at current stage the initial phase is not optimized when evaluating FF. Despite that, our best waveform models can be shown to achieve more than $97\%$ accuracy even without optimizing the initial phase. Once the initial phase is also optimized, the accuracy can be expected to be further enhanced.

Note that it turns out that the cVAE models yield comparable but lower accuracy than the cAE models. This is probably because the template data sets considered in this paper are parameterized by two mass parameters, which cause not much degeneracy in mapping the parameter space to the template space. The degeneracy here means that the different set of parameters may yield quite similar waveforms. Thus, the variational feature of latent space of cVAE models may not be needed for such kind of deterministic training set. Despite that,  when considering more complicated template sets with more source parameters, the variational feature could be helpful to disentangle the degeneracy, which may occur more often. In this respect, it is still interesting to consider the VAE type models as a preliminary study for the future work. To not digress the main theme of this work, from now on we will simply focus on the various models based on the cAE scheme in the main text. As a comparison, in Appendix we present the performance of cVAE models with the same schematic structures of the cAE models.


\section{Conditional Autoencoder Waveform Models} \label{schemes}

Based on the cAE scheme we can construct various waveform models by different arrangements of the encoders and decoders. Since the input data are separated into keys and normalized strains, their associated cAE can be arranged to share a common decoder or not. For either cases, we consider two waveform models which further differ by how the labels are conditioned. In total, we will consider four waveform models, and compare their performances. In this way, we can understand the relevance of different arrangements to the performance, so that such experiences could be helpful for further constructions. Below we first consider the models in which the keys and normalized strains do not share a common decoder, and then the models do. 

\begin{figure}[htpb]
12\resizebox{8.5cm}{!}{\includegraphics{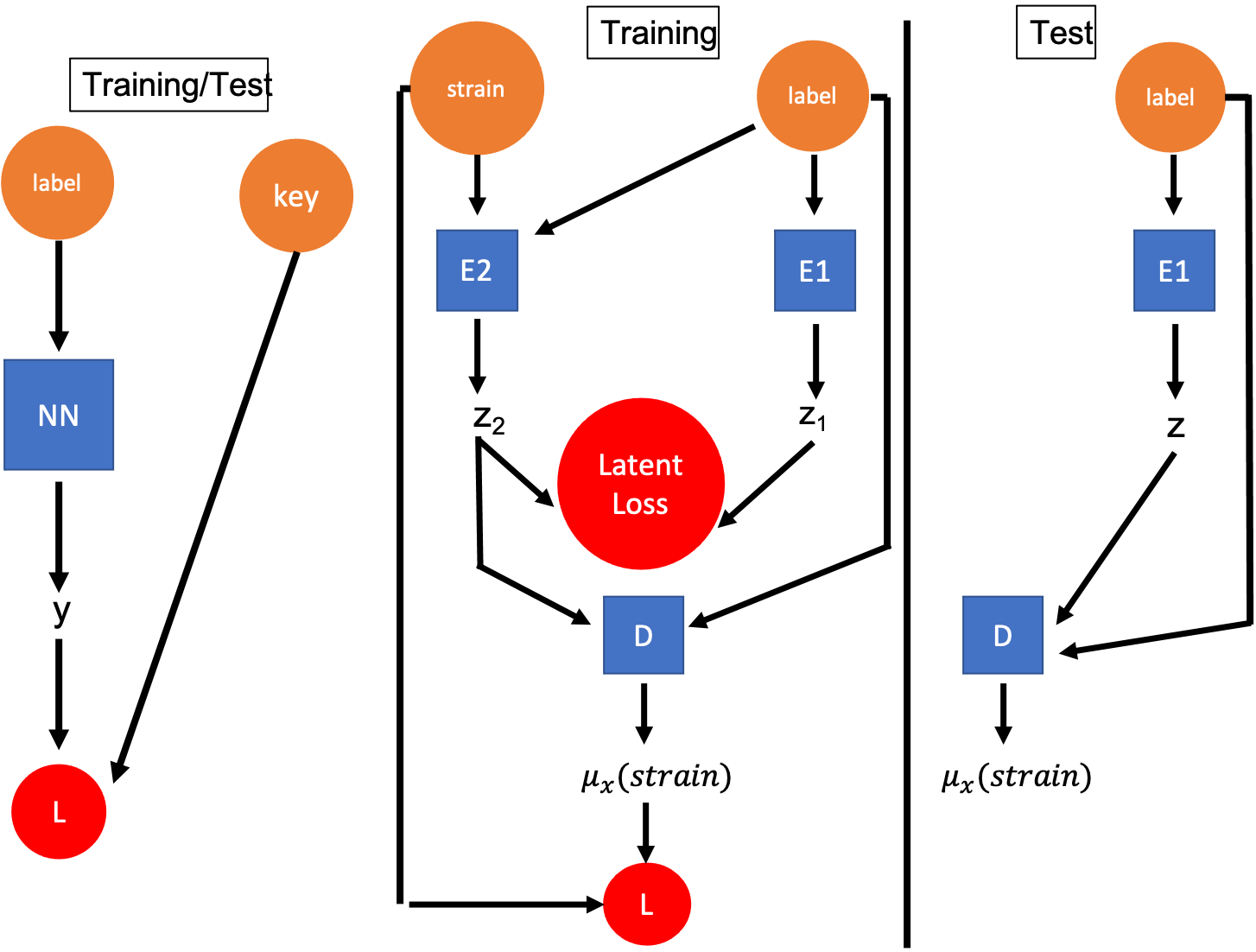}}
\caption{Schematic structure of the cAE+NN waveform model.}
\label{cAE+NN}
\end{figure}

\begin{figure}
\resizebox{8.5cm}{!}{\includegraphics{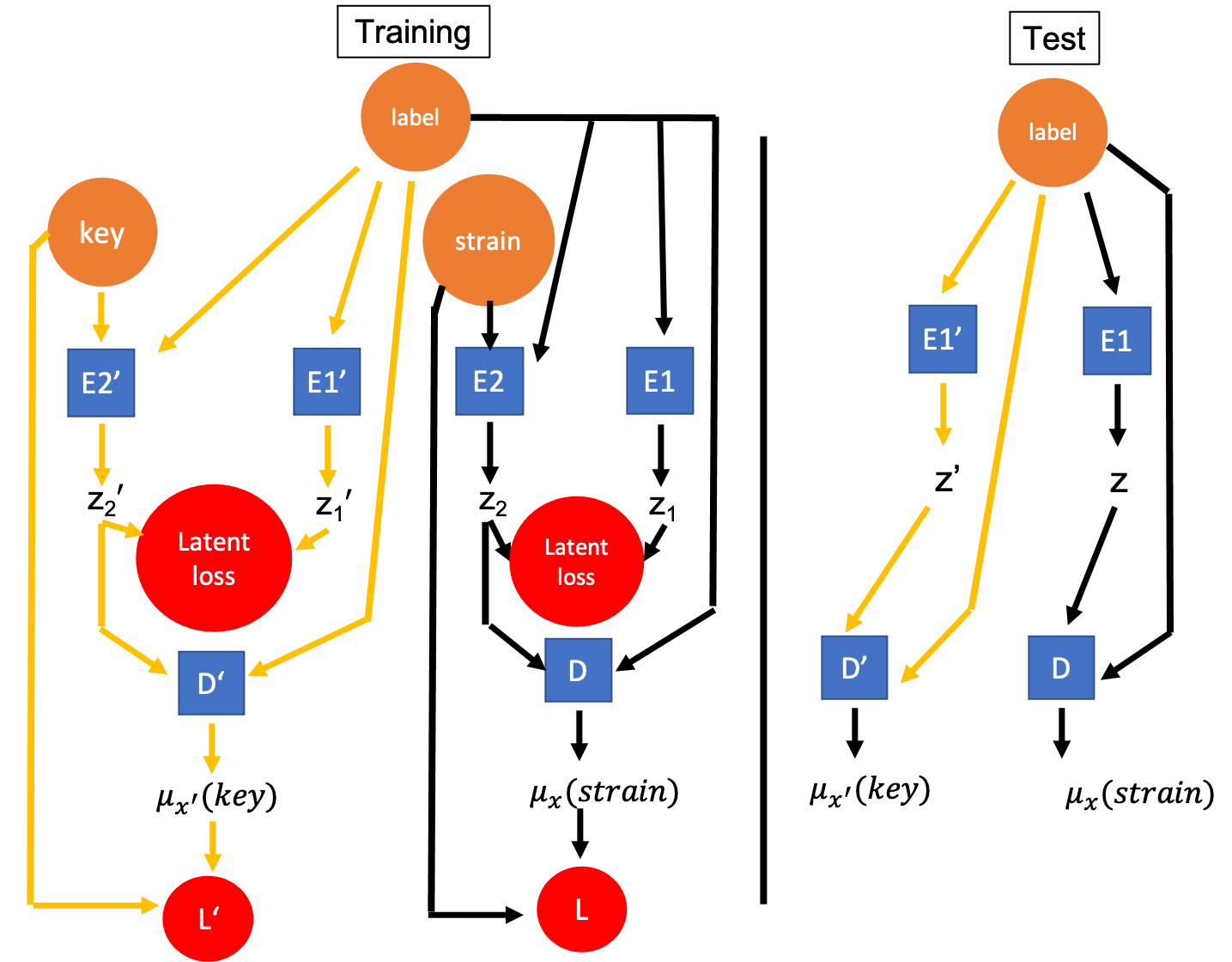}}
\caption{Schematic structure of the 2cAE waveform model. }
\label{2cAE}
\end{figure}

The first cAE waveform model as shown in Fig. \ref{cAE+NN} is what we call cAE+NN model, in which the strains are trained with cAE and the associated keys are trained with conventional supervised learning neural network (NN) because the dimensions of the key are relatively small. After done the training, we can drop the strain part from the model, and turn the remaining network (Right-most part of Fig. \ref{cAE+NN}) into a generative machine for waveforms. Since the keys and the normalized amplitude and frequency parts of the strains are trained separately, when generating a strain we need to combine them together to get the un-normalized amplitude $A(t)$ and frequency $\omega(t)$. Finally, we need to integrate the frequency to get the phase $\theta(t)$ and then combine with the amplitude  to get the strain $h_{\rm ML}(t)$. With the output strain $h_{\rm ML}(t)$ we can use \eq{accuracy-rate} to evaluate the FF for each waveform in test data set, and then take the average to obtain the accuracy. The resultant performance is $85.73\%$ for LMR, and $55.95\%$ for HMR. As expected, the accuracy for HMR is lower than the one for LMR. The accuracy is not good enough for the purpose of data analysis. This is because the simple NN for training the key is not the optimal scheme as AE.

By the faith on the power of semi-supervised training, we can replace the NN by AE to train the key, and the new scheme is shown in Fig. \ref{2cAE}. As expected, the performance for both LMR and HMR get improved. The resultant accuracy is $89.92\%$ for LMR and $67.20\%$ for HMR. The accuracy of LMR is now barely good for detection purpose, but not good for PE to extract the source properties. Still the HRM part is not accurate enough for practical purpose.

\begin{figure}
\resizebox{8.5cm}{!}{\includegraphics{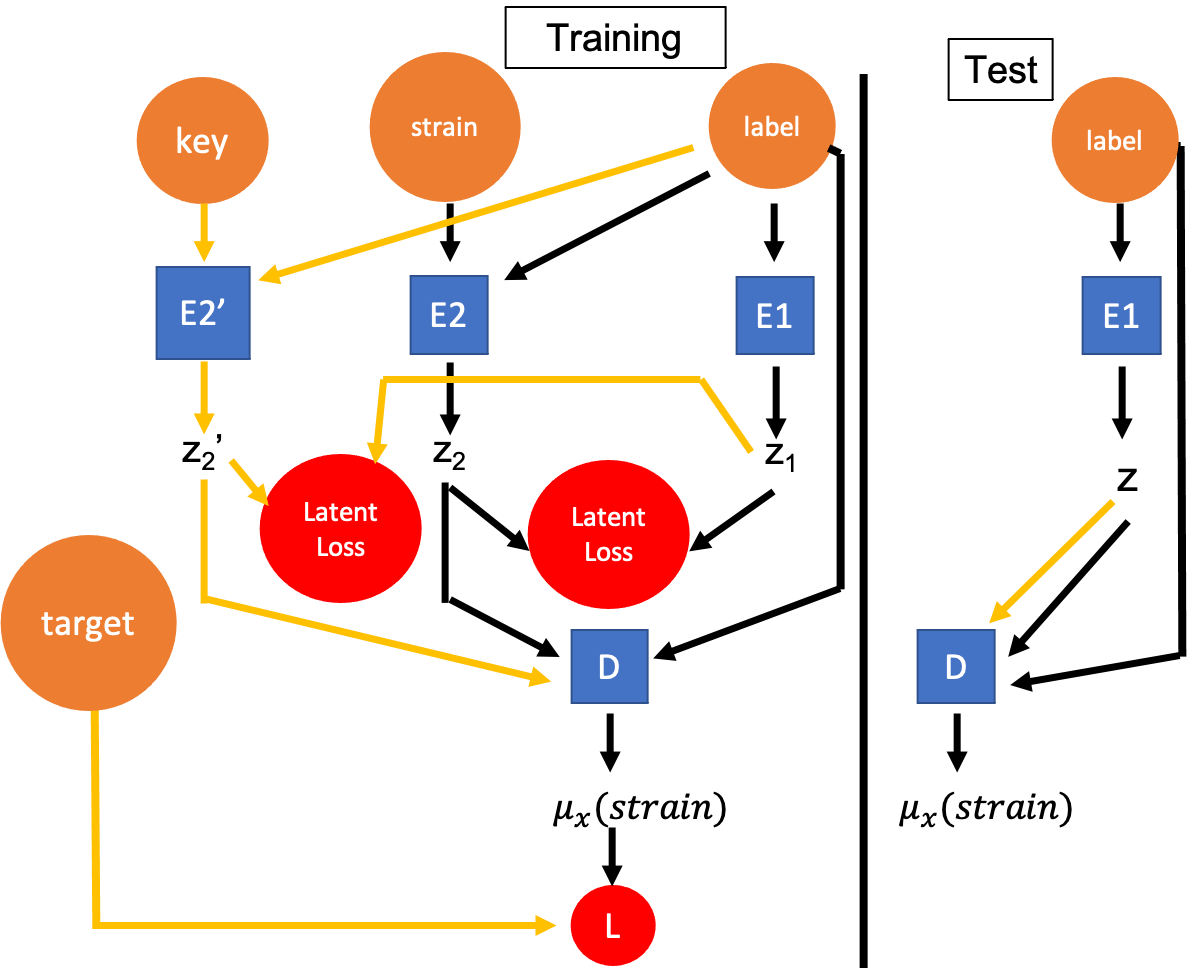}}
\caption{Schematic structure of the 1c2E1D waveform model.}
\label{1c2E1D}
\end{figure} 

\begin{figure}
\resizebox{8.5cm}{!}{\includegraphics{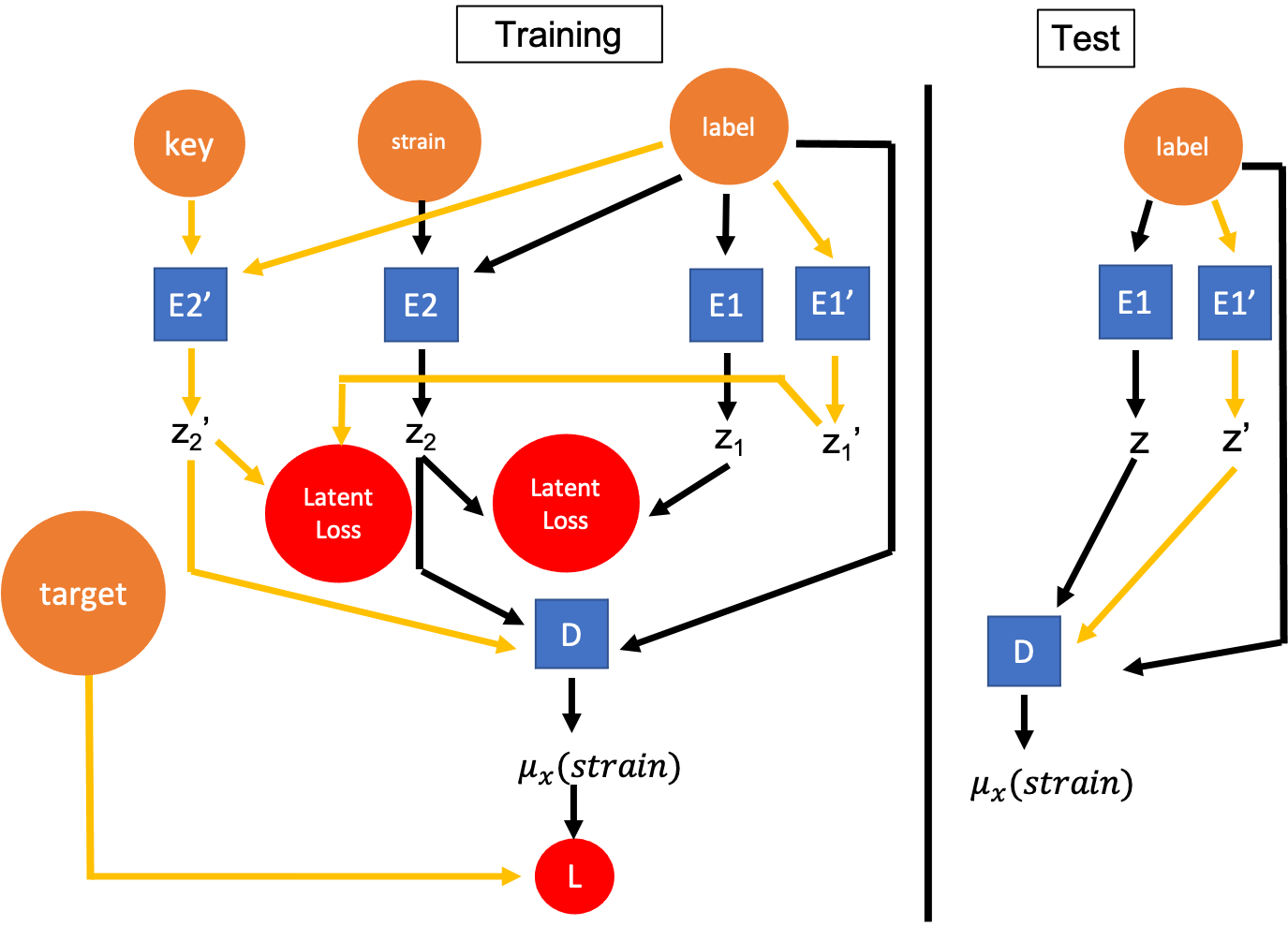}}
\caption{Schematic structure of the 2c2E1D waveform model.}
\label{2c2E1D}
\end{figure} 

A common feature of the above two models is that they train the keys and strains separately, and the correlation is only through the common labels. Instead, we can correlate the outputs of the different encoders into a common decoder, and directly compare the decoder's output to the corresponding un-normalized frequency and amplitude of the input strain to obtain the reconstruction loss. Intuitively, the additional correlation may improve the accuracy of the generative model.  We will consider two such kinds of models. The first one is shown in Fig. \ref{1c2E1D}, which we call 1c2E1D, i.e., one conditional encoder, and two waveform encoders (one for the strain and one for the key), and finally one common decoder. In this model, the latent sizes of all the encoders should be the same. This may cause redundancy of the latent space for training the key since the key's dimension is far less than the strain's. To remedy this, we introduce the second model as shown in Fig. \ref{2c2E1D}. We call this model 2c2E1D, i.e., now we have two conditional encoders so that the latent
sizes of the encoders for the strain and the key can be different. Moreover, since now the keys and the normalized strains share a common encoder, we can in fact choose the target in Fig. \ref{1c2E1D} and \ref{2c2E1D} to be the un-normalized amplitude and frequency, i.e., $A(t)$ and $\omega(t)$, but not the normalized $\hat{A}(t)$ and $\hat{\omega}(t)$, to evaluate the reconstruction loss.  For simplicity, we choose the MSE for the reconstruction loss so that the minimization of the reconstruction loss will make the machine to generate $\tilde{x}$ being as close to the target $A(t)$ and $\omega(t)$ as possible. In contrast to the cAE+NN and 2cAE waveform models, this will save the procedure of converting the normalized strains into their un-normalized counterparts.

\begin{table*}
	\centering
	\scalebox{1.2}{
	\begin{tabular}{|*{5}{c|}}
		\hline
		{} & cAE+NN & 2cAE & 1c2E1D & 2c2E1D \\
		\hline
		accuracy (LMR) & 85.73\% & 89.92\% & 97.65\% & 98.20\%\\
		\hline
		accuracy (HMR) & 55.95\% & 67.20\% & 97.70\% & 97.02\%\\
		\hline
	\end{tabular}}
	\caption{Summary of the accuracy of the low-mass-ratio (LMR) and high-mass-ratio (HMR) waveforms for each cAE waveform model considered in this paper. The accuracy is the average of the fitting factors (FFs) (see \eq{accuracy-rate}) for all the test data. We see that both 1c2E1D and 2c2E1D models can have accuracy more than $97\%$ for both LMR and HMR.}
	\label{summary acc 20}
\end{table*}

\begin{table*}
	\centering
	\scalebox{1.2}{
	\begin{tabular}{|*{3}{c|}}
		\hline
		FFs for 2c2E1D & LMR & HMR \\
		\hline
		Minimum FF & 82.49\% & 74.13\% \\
		\hline
		Median FF & 98.61\% & 98.22\% \\
		\hline
		Maximum FF & 100.0\% & 99.99\% \\
		\hline
	\end{tabular}}
	\caption{Summary of the FFs of the LMR and HMR waveforms for 2c2E1D cAE waveform model considered in this paper. The associated cumulative distribution function of FFs is shown in Fig. \ref{overlap distribution}. We see that the medians are comparable with accuracy listed in Table \ref{summary acc 20}.}
	\label{summary acc 20 med}
\end{table*}

\begin{table*}[htbp]
	\centering
	\scalebox{1.2}{
	\begin{tabular}{|*{5}{c|}}
		\hline
		{} & cAE+NN & 2cAE & 1c2E1D(cAE) & 2c2E1D(cAE)\\
		\hline
	Training time (strain)(sec) & 4042.5 & 3329.4 & \multirow{2}*{4536.6} & \multirow{2}*{4462.6}\\
		\cline{1-3}
	Training time (key)(sec) & 81.2 & 144.8 &  &\\
		\hline
		Generations/epochs (strain) & 8000 & 8000 & \multirow{2}*{10000} & \multirow{2}*{10000}\\
		\cline{1-3}
		Generations/epochs (key) & 10000 & 10000 & &\\
		\hline
	Generation time per waveform (milli sec) & \multicolumn{4}{|c|}{0.8-1.0} \\
		\hline	
	\end{tabular}}
	\caption{Summary of the training time, generation/epoch number and the generation time of a single waveform for each cAE waveform model considered in this paper. Note that it takes less than 1 millisecond to generate a single waveform. This is about 10 to 100 times faster than the EOB running on the same computing facility.}
	\label{summary speed 20}
\end{table*}

The resultant performance for the above two models are the following. For the 1c2E1D waveform model, the accuracy is $97.65\%$ for LMR, and $97.70\%$ for HMR. For the 2c2E1D waveform model, the accuracy is $98.20\%$ for LMR, and $97.02\%$ for HMR. Their performances are comparable, and are accurate enough for both LMR and HMR, i.e., greater than $97\%$. Note that we have not optimized the overlap accuracy over the initial phase yet, once it is done we will expect higher overlap accuracy \footnote{Our preliminary study shows that it can achieve almost $99\%$ for LMR and $98\%$ for HMR.}. The high accuracy of the generated waveforms indicates that both models are good for the purpose of low latency detection and for PE of gravitational wave events with improvement of accuracy in the future work. The high accuracy for HMR part can also be exploited for progressively self-training to generate waveforms of HMR.

We summarize in Table \ref{summary acc 20} the accuracy for each cAE waveform model considered in this paper. The 1c2E1D and 2c2E1D model are the best in the accuracy rate. Overall the 2c2E1D model is slightly superior. To characterize its detailed performance, we also give the minimum FF, median FF and maximum FF of this model in Table \ref{summary acc 20 med}, from which we see the median FF is comparable with the accuracy, i.e., the average of FFs. Later we will discuss more details for this model.  Note that, the above models are all implemented based on the cAE scheme.  We can also replace the cAE schemes in these models by the cVAE ones, and obtain the corresponding cVAE waveform models. However, there is one additional model called cVAE+cAE, see Fig. \ref{cVAE+cAE} in the Appendix, in which we use cAE to train the keys and cVAE to train the strains.  The performances of these cVAE waveform models are listed in the Table \ref{summary acc 20_VAE} of the Appendix. It turns out that the accuracy of the cVAE waveform models are comparable to their cAE counterparts. However, by examining in more details it seems that cAE models are superior than the cVAE ones, even for the HMR. This is a bit surprising that the generative nature of VAE does not help to improve the accuracy.

Besides, we also summarize in Table \ref{summary speed 20} the training time and generation/epoch number of the waveform models and the generation time of a single waveform  for each cAE waveform model considered in this paper. 
The run-times of the cVAE waveform models are comparable with their cAE counterparts listed in Table \ref{summary speed 20}, thus are omitted for simplicity. We see that the training time is about $4000$ seconds for all the waveform models, it is quite modest and implies that the extension to the full waveform models with more source parameters is manageable in the near future. Furthermore, the generation time of a single waveform is about one millisecond. Compared to the typical generation time for a EOB waveform, which is about few hundredths to few tenths of a second on the same computing facility, the speed enhancement is about $10$ to $100$ times.

\begin{figure} [htpb]
\resizebox{\hsize}{!}{\includegraphics{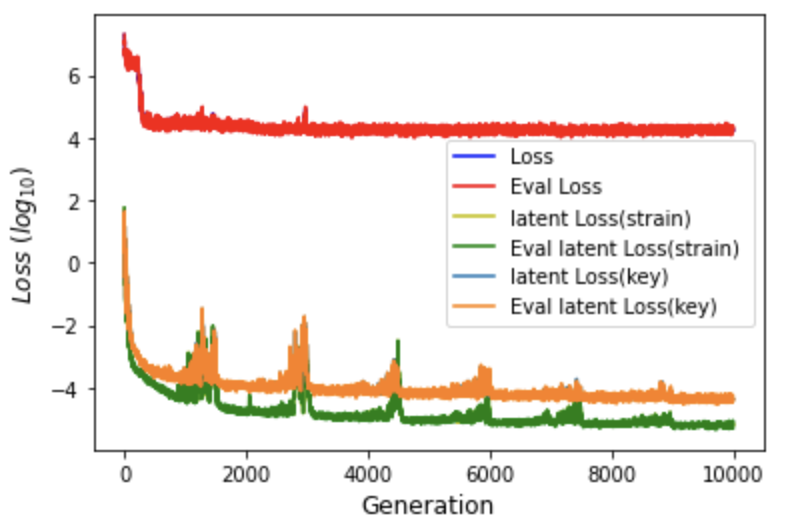}}
\caption{Histogram of the training losses for the 2c2E1D cAE waveform model. The types of training/validation losses are denoted by  (Loss/Eval$\_$loss, latent Loss(strain)/Eval latent Loss(strain), latent Loss(key)/Eval latent Loss(key)) in the graphic illustrations, which literally mean the total loss, the latent loss of the normalized strain, and the latent loss of the key, respectively. The match of Loss and latent Loss implies no over-fitting. The overall trends show that the training is stabilized around $10000$ generations/epochs.}
\label{training losses}
\end{figure}

As the 2c2E1D waveform model is the best among all the waveform models considered in this paper, we look into some details of this model. First, we list the hyperparameters of this model in in Table \ref{parameters 2c2E1D}, and histogram of its training losses in Fig. \ref{training losses}. Based on the information one can reproduce the model quite easily. From Fig. \ref{training losses} we see that the training and validation losses match well and stop increasing around $8000$ generations/epochs. This implies that our training is not over-fitted and stabilized.

\begin{table}[htbp]
	\centering
	\scalebox{1.1}{
	\begin{tabular}{|*{6}{c|}}
		\hline
		{} & $E_1$ & $E_2$ & $E_1^{'}$ & $E_2^{'}$ & Decoder \\
		\hline
		Latent size & 8 & 8 & 3 & 3 & 8 \& 3\\
		\hline
		CNN layers & \textcolor{gray}{None} & 2 & \textcolor{gray}{None} & \textcolor{gray}{None} & 3\\
		\hline
		Filter size & \textcolor{gray}{None} & [16,16] & \textcolor{gray}{None} & \textcolor{gray}{None} &[4,4,4]\\
		\hline
		conv features & \textcolor{gray}{None} & [5,15] & \textcolor{gray}{None} & \textcolor{gray}{None} & [16,32,64]\\
		\hline
		Pool size & \textcolor{gray}{None} & [4,4] & \textcolor{gray}{None} & \textcolor{gray}{None} & [4,4,4]\\
		\hline
		dilation rate & \textcolor{gray}{None} & \textcolor{gray}{None} & \textcolor{gray}{None} & \textcolor{gray}{None} & [1,2,2]\\
		\hline
		NN layers & 4 & 3 & 4 & 3 & 6\\
		\hline
		Neural size & 500 & 500 & 400 & 400 & 800\\
		\hline
	\end{tabular}}
	\caption{Hyperparameters for the 2c2E1D model. Here ``conv features" is the abbreviation of the convolution features.}
	\label{parameters 2c2E1D}
\end{table}

\begin{figure}[htpb]
\resizebox{\hsize}{!}{\includegraphics{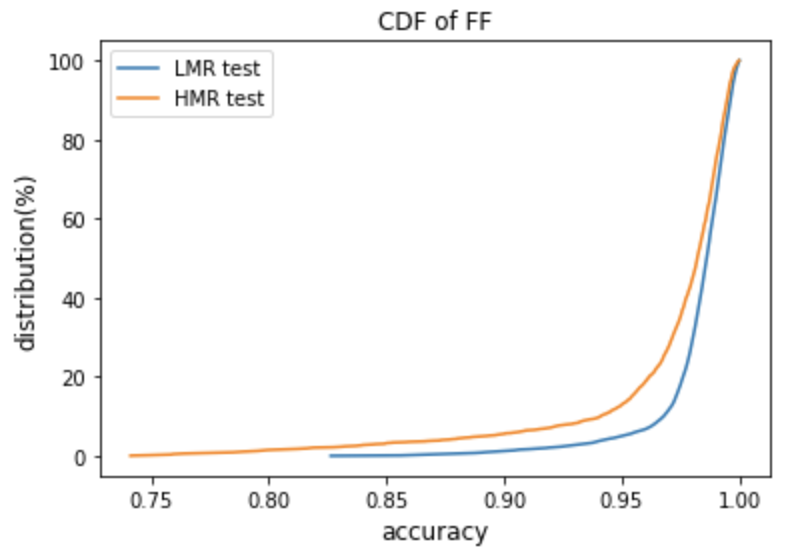}}
\caption{Cumulative distributions functions of fitting factors (FFs) of 2c2E1D cAE waveform model for the LMR (blue) and HMR (orange) generated waveforms. As expected, the HMR one has a broader tail. Overall, the outliers are rare.}
\label{FF tomography}
\end{figure}  

\begin{figure}[htpb]
\resizebox{\hsize}{!}{\includegraphics{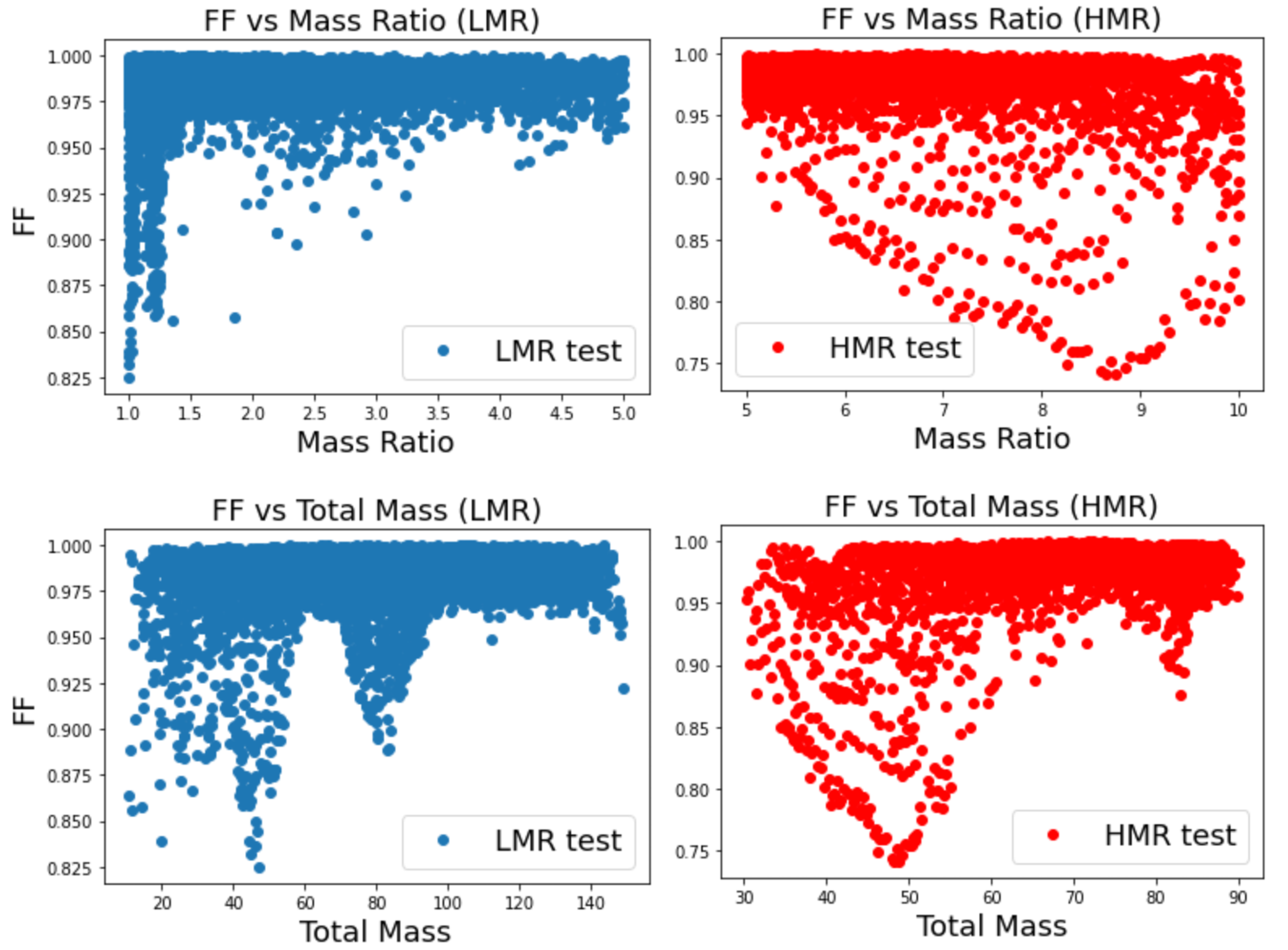}}
\caption{Distributions of FFs of 2c1E1D cAE waveform model as functions of mass ratio (Upper row) and total mass (Lower row) for LMR (blue) and HMR (red) test data set.  Note that each dot represent one template in the test data set.}
\label{overlap distribution}
\end{figure}

Further, we can understand more the tomography of the accuracy for the 2c2E1D cAE waveform model by plotting the cumulative distributions function (CDF) of the FF for both LMR and HMR. The results are shown in Fig.\ref{FF tomography}. As expected, we see that the HMR one has a broader tail, however,overall the FFs are concentrated on the the side of high FF near $100\%$. This implies the outliers are rare, and the generated waveforms can be reliably implemented for the practical data analysis such as the detection and PE of the gravitational wave events. For curiosity, it is also interesting to see how FF changes with the mass ratio and total mass of the BBH. We present the distributions for 2c2E1D cAE waveform model in Fig. \ref{overlap distribution}. We see that the low FFs appear more often in the regime of lower mass ratio and smaller total mass for LMR ones. The latter one could be due to the 
the fact that the templates in this regime are less loud (small mass), but it is not clear about the cause for the former one (lower mass ratio). On the other hand, the FF for the HMR does not behaves the same way, which could be due to the insufficient training data.

Finally, the typical neural network structure of the above cAE models is given in Fig. \ref{machine structure} in Appendix.

\section{Conclusion and Discussion} \label{conclusion}
In this paper we construct various waveform models based on the neural network of conditional autoencoders (cAE) and its variational extensions (cVAE). Their accuracy and run-time have been summarized in Table \ref{summary acc 20} and Table \ref{summary speed 20} in the main text for cAE models, and in Table \ref{summary acc 20_VAE} in Appendix for cVAE models, respectively. For simplicity, we represent the accuracy of these generative waveform models by the the average of fitting factor, which is based on the waveform overlap of the matched filtering. Among these waveform models, the so-called 2c2E1D cAE model is the best with more than $97\%$ for both the low-mass-ratio (LMR) and high-mass-ratio (HMR) waveform generation. This demonstrates the viability of our best waveform model to be implemented in the practical gravitational wave data analysis and parameter estimation (PE). Especially, the generation time of a single waveform is $10$ to $100$ times faster than the traditional EOBNR method, it implies that the waveform generation for the low latency detection can be accelerated by our waveform models. With the improvement of the accuracy in the future work, the revised version of our generative waveform model may also help to speed the parameter estimation. 
 
Moreover, the impressive accuracy for HMR waveform generations is encouraging because fraction of the HMR waveforms in the training and validation data set is less than $3\%$. This implies that one may be able to generate higher mass-ratio waveforms by a series of self-training with the generative outputs of the lower-mass ratio machine as the training data for the higher mass-ratio ones. This may open a new venue to generate the waveforms with intermediate mass-ratio, say greater than $15$.

Despite that, there are still ample space to improve our waveform models. As a proof-of-concept study, we only consider the inspiral-merger part of the full waveforms. Although the ringdown part is quite short but it contains the information of quasi-normal modes. We are currently training the waveform models for the full waveform based on the similar cAE or cVAE scheme, and will report our results in the near future. Moreover, to be more useful in the practical data analysis tasks, we shall also include more source parameters such as spins, precession and tidal deformabilities. Once the above goals are achieved, we can incorporate our waveform models to the standard pipeline of detection and PE, and help to accelerate the data analysis tasks in the coming O4 and O5 observation runs of LIGO/Virgo/KAGRA.

Added-in-proof. When finalizing the draft, an eprint \cite{Lee:2021isa} with the similar goal appears, in which a RNN framework is adopted to generate the merger-ringdown parts of the waveform from the input associated inspiral one.

\section*{Acknowledgement}
We thank Kai-Feng Chen, Yao-Yu (Joshua) Lin and members of TGWG and CAG for discussions and comments. In particular, we also thank Han-Shiang Kuo for the discussions on cVAE structures, and Jie-Shiun Tsao for helping the sever implementation. CHL would like to thank Wei-Ren Xu for his guidance into the machine learning. This work is supported by Taiwan's Ministry of Science and Technology (MoST) through Grant No.~109-2112-M-003-007-MY3.  We also thank NCTS for partial financial support.

\appendix
\section*{Appendix: Structures and Results of cVAE waveform models}
\begin{figure}[htbp]
\resizebox{8cm}{!}{\includegraphics{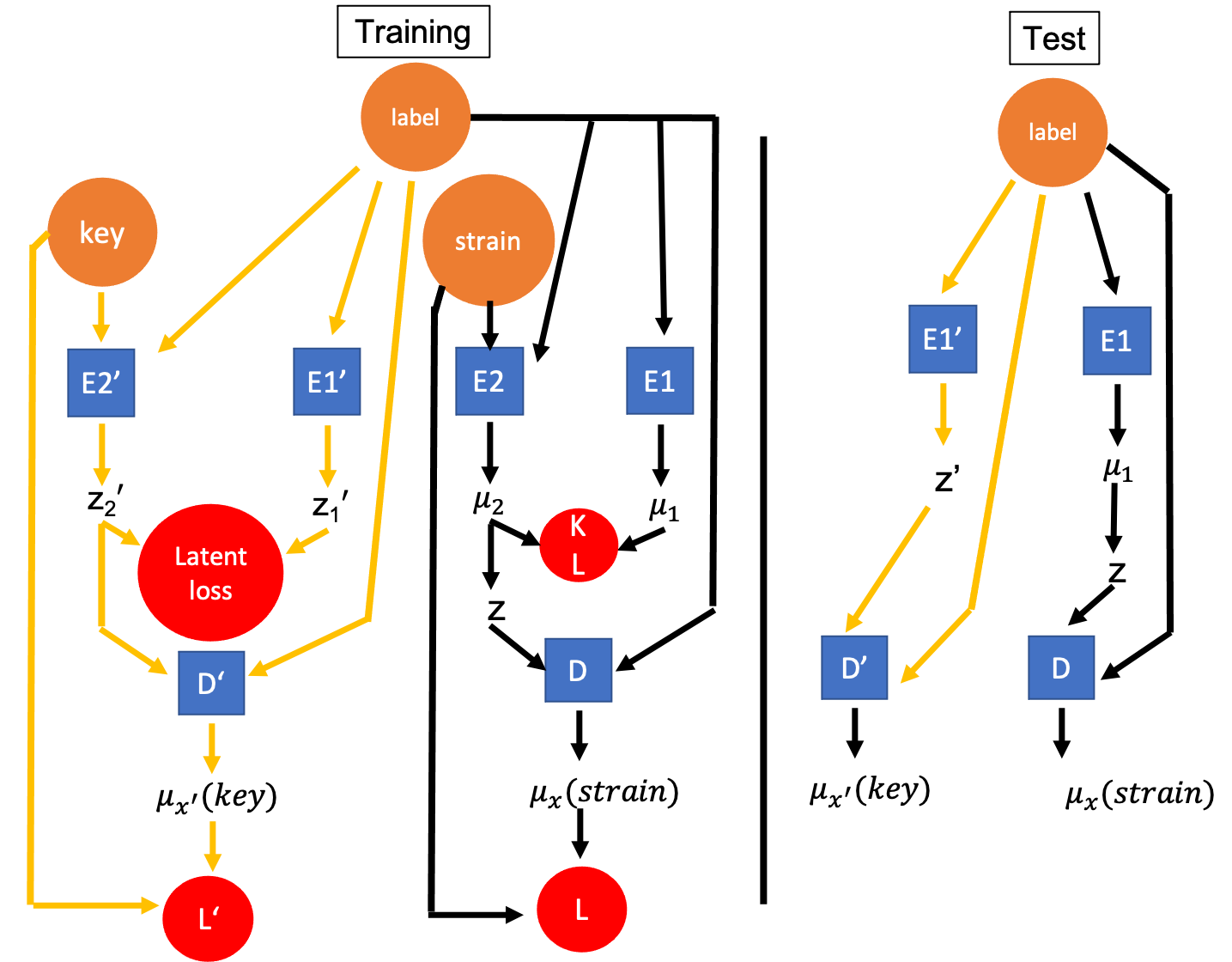}}
\caption{Schematic structure of the cVAE+cAE waveform model. }
\label{cVAE+cAE}
\end{figure}

\begin{figure*}[htbp]
\begin{minipage}[htbp][11cm][t]{0.9\textwidth}
\centering\resizebox{18cm}{!}{\includegraphics{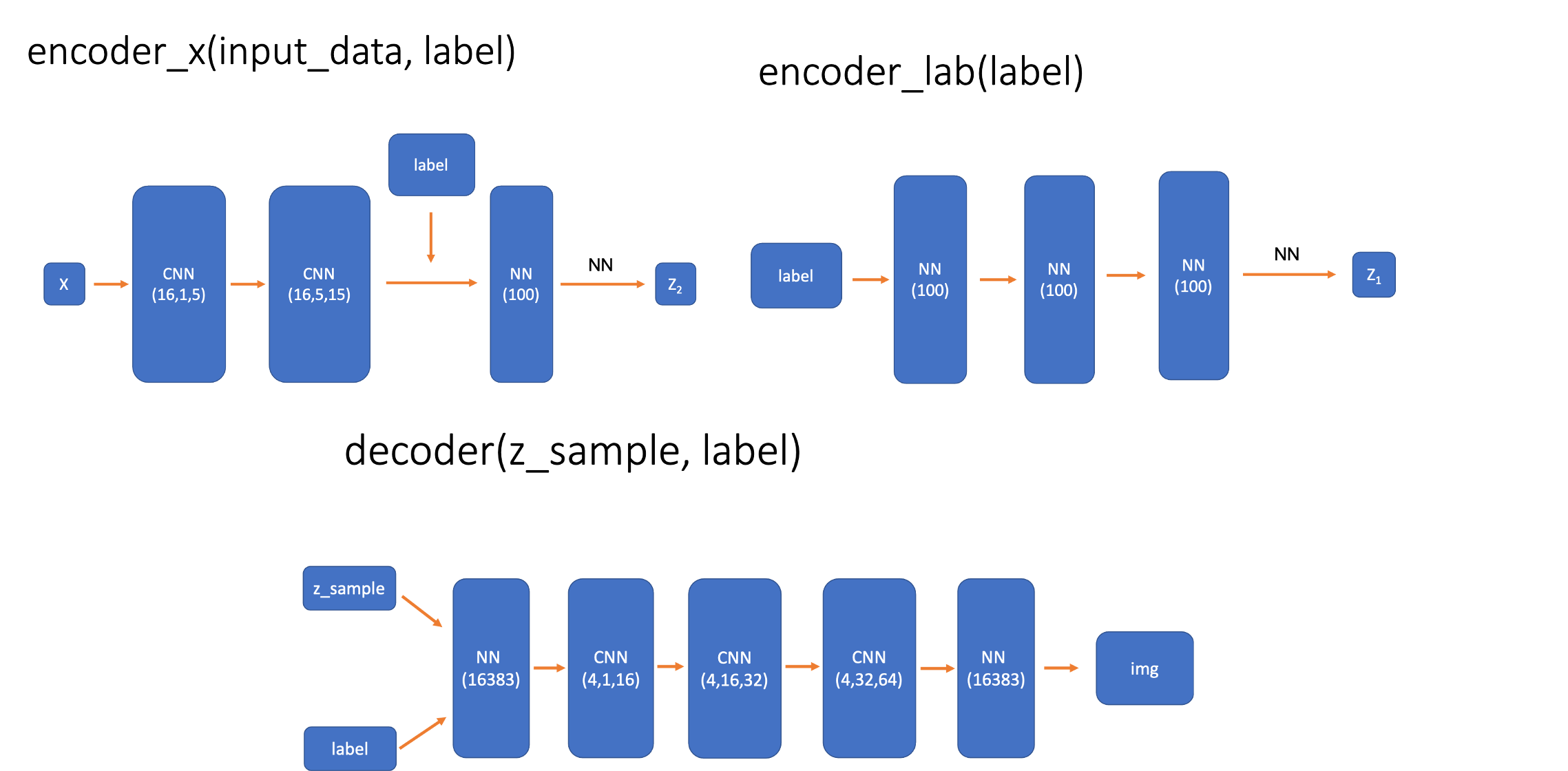}}
\caption{A typical machine structure with details of hyperparameters for the cAE or cVAE used in this work. Each neural network (NN) or convolutional neural network (CNN) is denoted by a box with its dimension specified. {\bf Top-left:} the encoder for the strain. {\bf Top-right:} the encoder for the label. {\bf Bottom:} the decoder to reproduce the strain by inputting the latent vector and the label.}
\label{machine structure}
\end{minipage} 
\end{figure*}

In this Appendix, we summarize the performance of the cVAE counterpart of the cAE waveform models considered in the main text. These counterpart models are simply obtained by replacing the cAE with cVAE in the associated cAE waveform model. However, for the 2cAE model, we can in fact replace only the cAE for the strains by cVAE, and still keep the cAE for the keys intact. In this way, we have a new model called cVAE+cAE model as shown in Fig. \ref{cVAE+cAE}. For all the cAE and cVAE waveforms models considered in this work, the typical neural network structure is shown in Fig. \ref{machine structure}, which serve as the guideline for the readers to implement the coding.

\FloatBarrier

The accuracy for the cVAE waveform models are shown in Table \ref{summary acc 20_VAE}, from which one can compare with Table \ref{summary acc 20} and finds that the accuracy are comparable for the cVAE models and their counterparts.  Besides, the run-times for these cVAE models are comparable with their cAE counterparts, i.e., about $4000$ seconds for training and $1$ milli second to generate a single waveform, thus for simplicity we will not listed here.  Finally, one more point is that after training is done, we will just choose the mean value of the latent layer to generate the waveform to avoid the stochastic feature of VAE.

\begin{table}[ht]
\begin{minipage}[htbp][5cm][t]{0.45\textwidth}
	\centering
	\scalebox{0.95}{
	\begin{tabular}{|*{6}{c|}}
		\hline
		{} & cVAE+NN & cVAE+cAE & 2cVAE & 1c2E1D & 2c2E1D\\
		\hline
		accuracy & \multirow{2}*{89.73\%} & \multirow{2}*{89.03\%} & \multirow{2}*{73.23\%} & \multirow{2}*{94.35\%} & \multirow{2}*{97.16\%} \\
	    (LMR) & & & & &\\
		\hline
		accuracy & \multirow{2}*{65.56\%} & \multirow{2}*{70.26\%} & \multirow{2}*{73.65\%} & \multirow{2}*{79.11\%} & \multirow{2}*{91.92\%} \\
		(HMR) & & & & &\\
		\hline
	\end{tabular}}
	\caption{Summary of the accuracy for both LMR and HMR for  the cVAE counterpart of each cAE waveform model considered in the main text.}
	\label{summary acc 20_VAE}
\end{minipage}
\end{table}

\bibliographystyle{abbrv} 
\bibliography{cVAE.bib} 

\end{document}